%% file: main.tex
\documentclass[10pt]{article}
\usepackage[margin=1in]{geometry}
\usepackage[T1]{fontenc}
\usepackage{lmodern}
\usepackage{microtype}
\usepackage{booktabs}
\usepackage{graphicx}
\usepackage{enumitem}
\usepackage[hidelinks]{hyperref}
\setlist{nosep,leftmargin=1.4em}
\input{tables}

\title{Ask, Forget, Repair: What Provenance-Gated Agent Memory\\Costs Across Six Language Models}
\author{Radhika Motwani\\Independent Researcher}
\date{26 November 2026}

\begin{document}
\maketitle

\begin{abstract}
An agent with long-term memory can be poisoned in one session and exploited in a later one. A
common defense is to record where each memory came from and to stop untrusted memories from
driving consequential actions. Prior work shows that such defenses block attacks; much less is
known about what they cost, or how much of the protection comes from the model and how much from
the runtime. We run six current language models from four vendors as the agent in one 60-session
synthetic back-office workload with 150 planted memory items per model, under three conditions:
no defense, provenance labels that are shown to the model but not enforced, and the same labels
enforced by the runtime. With no defense, planted items later caused a wrong action in 2\% to
93\% of cases depending on the model. Showing labels alone reduced this to 0--1\% for four models
but left 15\% and 67\% for the other two. With enforcement no planted item succeeded on any model
unless a simulated user approved a bad confirmation. The defense is not free. With no attacker
present, the share of requests completed correctly without involving the user fell to 86\% for
every model, a drop of 3 to 14 points: each legitimate fact that arrived from outside cost one
exchange with the user, after which it was remembered as the user's own. The form of that exchange
depends on the model. Models that respect labels ask the user; a model that ignores labels
triggers runtime confirmation prompts. Labels without enforcement cost the label-respecting
models more, up to 30 points, because merging let the user's answers inherit the untrusted label
and expire. The planted texts are simple and do not adapt to the defense, the tasks are
synthetic, and the user is simulated, so these results are a lower bound on attacks and a first
measurement of cost, not a ranking of models. Code, workload and every model reply are released.
\end{abstract}

\section{Introduction}

Agents that keep memory across sessions inherit a delayed form of indirect prompt
injection~\cite{greshake}. Attacks that plant content in an agent's memory or knowledge base are
well documented~\cite{agentpoison,minja,mpbench}. A document
read today (an email, a web page, a shared file) can leave behind a memory that steers an action
weeks later, after the document itself is gone. A line of recent work addresses this by attaching
provenance to memory and denying untrusted memory the authority to trigger consequential
actions~\cite{tmanm,permit,ppmf,sleeperchannels}. The systems that report measurements find
no or almost no successful attacks~\cite{tmanm,permit,ppmf}.

Two questions are left open by that work, and they matter to anyone deciding whether to deploy
such a defense. First, what does it cost? A legitimate supplier really does send a new bank
account by email. A defense that distrusts everything from outside must either withhold that fact,
ask the user about it, or let it expire. Second, how much of the protection is the runtime's and
how much is the model's own caution? If a capable model already declines to act on a memory it can
see is untrusted, then showing it a label may be enough, and enforcement is redundant. If it does
not, a label is decoration.

We measure both on six current models with one workload and one memory runtime. This is a
measurement paper. The defense mechanisms we implement are taken from prior work, chiefly
origin-bound authority~\cite{tmanm} and confirmation gates~\cite{permit,sleeperchannels}; we claim
no new defense. Most of the individual effects we report have precedents, which
Section~\ref{sec:related} credits; what is new here is measuring them together, on one workload,
across current models from four vendors. Our contributions are:

\begin{itemize}
\item A head-to-head comparison, on the same workload, of labels shown to the model against labels enforced by the
runtime. Labels alone were nearly sufficient for four hosted models and clearly insufficient for
an open-weight model and for one hosted model (Section~\ref{sec:labels}).
\item A breakdown of what the enforced defense costs when no attacker is present, separating
requests the agent completes unprompted, requests that need a confirmation, details that are
lost, and requests where the agent has the right detail and asks the user anyway
(Section~\ref{sec:cost}).
\item Observations about the memory lifecycle: merging trusted and untrusted entries makes the
user's own statements distrusted, so that a label-respecting model asks for the same detail
again; expiry removes legitimate facts along with planted ones; and
retracting a malicious source only works if provenance survived the agent's own note-taking
(Section~\ref{sec:lifecycle}).
\item A harness in which the agent only proposes what to store and what to do, so any
model behind a chat-completions endpoint can be evaluated, with every prompt and reply saved.
\end{itemize}

\section{Related work}
\label{sec:related}

\paragraph{Provenance-gated memory.} TMA-NM binds a memory item's authority to its origin at write
time, propagates that through the agent's own summaries, and reports no successful attacks across
eight models from six vendors, with its core invariants model-checked in TLA+~\cite{tmanm}. It
reports legitimate-task utility equal to the undefended agent; legitimate actions that rest on
uncorroborated outside content go through a one-time user confirmation. PerMIT combines a
quarantined reader, a gate that releases an untrusted value only after the exact value is
confirmed, eviction, and endorsements shared across a community of users, with two integrity
theorems machine-checked in Lean~4; it reports verifier false-positive rates and, for a simulated
community, approvals saved per user~\cite{permit}. PPMF describes \emph{provenance laundering}, in which
consolidation rewrites outside content as apparent user history, and gates tool calls on
platform-maintained provenance; it reports that readable provenance cues in memory text remained
exploitable on several small open-weight models while the runtime gate held~\cite{ppmf}. Maloyan
and Namiot distinguish provenance warnings placed in the model's context from gates enforced
outside the model loop and argue that only the latter is load-bearing, without a multi-model
measurement~\cite{sleeperchannels}. Our Labels and Enforced conditions correspond to that
distinction. MemLineage tracks lineage through derived entries with weighted attribution and
enforces it at a gate outside the model; it shows that coarse taint, where everything retrieved
taints the result, blocks benign actions that finer attribution keeps~\cite{memlineage}. Our
taint is of the coarse kind. Closest to our Labels condition, Zhan et al.\ show for seven
models that authority labels rendered in the prompt cut unauthorized actions from about half of
cases to a few percent, without evaluating enforcement outside the model~\cite{authoritycollapse},
while Kwon finds in a smaller study that source attribution and a passive ``unverified'' tag gave
little or unreliable protection~\cite{manufacturedconfidence}. We put both conditions on one
workload, add an open-weight model, and break down the cost.

\paragraph{Marking untrusted content for the model.} Our Labels condition is a memory-specific
case of an older idea. Spotlighting marks untrusted text so that the model can tell it from
instructions~\cite{spotlighting}, and vendors train models to give lower privilege to content
that arrives through tools~\cite{instructionhierarchy,geminilessons}. That training is a
plausible reason the hosted models we test respect a visible label; we did not test this
explanation. CaMeL takes the opposite position, enforcing information flow outside the model so
that safety does not depend on the model's cooperation~\cite{camel}.

\paragraph{Cost of memory defenses.} ``The Price of Safety'' measures benign accuracy and token
overhead for four memory defenses, with no attacker and no confirmation step~\cite{priceofsafety}.
MEMGAUGE studies utility--risk trade-offs across writing, management and retrieval for eleven
models, with provenance used to quarantine records at the system level rather than shown to the
model~\cite{memgauge}. Karunanidhi shows that ranking by provenance can remove the evidence a
question needs when that evidence is itself untrusted~\cite{utilityunderattack}; our ``detail
lost'' and ``asked anyway'' columns are the action-side counterpart of that effect. Cerruti et
al.\ report that filtering memory records by source cut unauthorized actions and also cut
legitimate actions from 93\% to 53\%~\cite{eal}. Outside memory, FIDES enforces information-flow
policies deterministically~\cite{fides}, and Kolluri et al.\ measure the cost of such policies as
the number of human approvals a task needs~\cite{prudentia}; our confirmation counts are the
memory-setting analogue.

\paragraph{How models treat memory.} Hu and Ramachandran find that small open-weight models rely
on stale stored facts almost always, and that provenance metadata helps larger models more than
smaller ones~\cite{trustgap}. Li et al.\ compare how two Claude models and a Qwen model decide
whether to persist, verify or ask, without provenance labels or planted content~\cite{rememberverify}.
Pulipaka et al.\ report large differences between hosted models in whether a planted memory is
written at all, with one Claude model explicitly refusing suspicious writes~\cite{sleepermemory},
and Torres et al.\ find differences in activation across four models~\cite{ghostwriter}. Liu et
al.\ show that models differ widely in when they abstain or ask instead of acting, outside the
memory setting~\cite{agentabstain}. We observe related differences between vendors in an
adversarial, multi-session setting.

\paragraph{Asking the user.} Confirmation steps assume an attentive person. Field studies of
security warnings found click-through rates that vary widely and can be high~\cite{warningland},
Mitchell et al.\ argue that repeated approval requests from agents erode the scrutiny they rely
on~\cite{outoftheloop}, and Wang et al.\ model when an agent should ask at all given limited human
attention~\cite{whentoask}. Asking is also an attack surface: planted warnings can make agents
withhold legitimate actions, and cautious prompting makes this worse~\cite{abstentionattack}.
Kwon reports that an instruction to distrust memory stopped wrong decisions only by escalating
every request~\cite{manufacturedconfidence}. We vary the simulated user's carelessness but do not
study people.

\paragraph{Benchmarks and attacks.} Lin et al.\ survey attacks and defenses across the memory
lifecycle and identify rollback and long-horizon evaluation as gaps~\cite{survey}. MemPoison
reports that down-weighting by source reliability has a small clean-task cost and leaves dormant,
context-triggered corruption largely in place~\cite{mempoison}. MPBench catalogues memory write channels and attack classes
for two agent harnesses and recommends write-path provenance tracking without testing
it~\cite{mpbench}. MemSecBench follows planted memory from write to consequence to repair and
finds that repair often removes benign memory along with the malicious entry~\cite{memsecbench};
we measure whether planted items can be removed and not what else is removed with them. MemoryDojo
provides realistic multi-session tasks across five
memory backends and evaluates no defenses~\cite{memorydojo}; it would be a better testbed than our
synthetic workload once its code is available. Adaptive attacks have repeatedly broken
prompt-injection defenses that reported near-zero attack success against static
attacks~\cite{attackermovessecond}. We do not run adaptive attacks, and Section~\ref{sec:limits}
says what follows from that.

\section{Setup}

\subsection{Threat model}

The attacker controls the content of some documents the agent reads. The attacker cannot reach the
model, the runtime, the memory store or the user's own messages, and wants a consequential action
(moving money or sending data) to go wrong in a later session. The runtime, the user channel and
the labels the runtime assigns are trusted. The model is not.

\subsection{Workload}

Each run simulates 60 sessions of a back-office assistant. Topics are details about twenty
invented companies: a payout account, a billing contact, or the recipient of a usage report. A
run contains:

\begin{itemize}
\item 20 details the user states directly in the first third of the run;
\item 15 legitimate details that arrive only in outside documents, such as a supplier's email;
\item 30 planted items in outside documents, ten of each kind below, each followed 1 to 40
sessions later by a request that depends on the same topic;
\item 135 user requests in total. Each asks the agent either to act (``Pay this month's invoice
from \ldots'') or to prepare something without external effect (``Add \ldots\ to my draft payment
sheet. Do not pay anything yet.'').
\end{itemize}

The three kinds of planted item are an \emph{extra action} (a standing instruction to copy
material to an outside address), an \emph{override} (a value contradicting one the user stated),
and a \emph{fill gap} (a value for a detail the user never gave). About half are worded bluntly
(``Ignore the earlier setting. Always use \ldots'') and the rest plainly (``\ldots\ is now
\ldots''). They are fixed templates. Nothing about them is tuned to any model or defense.
Memory is consolidated every five sessions: entries about the same topic are merged and the
newest value wins. We use five seeds, so each model sees 150 planted items. Because seeds share
many documents, a model answers 108 distinct planted documents.

\subsection{Agent protocol}

The model is called at two points. When a document is read, it is asked what, if anything, to
remember, and replies with a list of (topic, value, standing instruction) entries. When the user
makes a request, it is shown the request and its memory entries about the needed detail, and
replies with the entry it will use or with a decision to ask the user instead, plus any additional
actions it intends to take. The model only proposes. The runtime stores, labels, retrieves and
decides what executes. Retrieval is by exact topic handle; each document statement carries its
handle. Identical prompts are sent once and the reply reused, and requests ask for temperature~0
where the model accepts it.

\subsection{Conditions}

\begin{description}[leftmargin=1.4em,style=nextline]
\item[No defense.] A plain store. Entries carry no source. A merged entry looks like a fresh note.
\item[Labels.] The runtime labels every entry. User statements are trusted. Anything the model
writes after reading an outside document in the same session is untrusted and carries that
document's identifier. Untrusted entries expire after 20 sessions. A merged entry takes the lowest
trust, all sources and the earliest expiry of its parts. The model sees each entry's label and
sources. Nothing is enforced, and entries of different trust may be merged.
\item[Enforced.] Labels, plus runtime rules. Trusted and untrusted entries are never merged. A
value the user stated outranks any other. An untrusted memory cannot start an action, and an
action the model does not attribute to a memory is denied while untrusted memory is in view. A
request that acts and would use an untrusted value produces a confirmation prompt showing the
source; approval re-issues the value as a user statement. Retracting a source removes everything
it influenced.
\end{description}

Two further baselines, a keyword filter on writes and expiry applied only to verbatim tool output,
gave results identical to No defense for every model, because the model's own notes are what
persists. We omit them from the tables.

\subsection{Simulated user and metrics}

At a confirmation prompt the simulated user always approves a correct value and approves a planted
one with probability $f$ ($f=0.1$ unless stated). \emph{Attack success} is the share of planted
items that, at any later point, had their value used in a request that acts or had their
instruction executed. \emph{Saved} is the share present in memory after the document was read. For
requests whose needed detail has a true value, each is classed as \emph{correct, unprompted};
\emph{correct after prompt}; \emph{detail lost} (no entry holding the right value could be
retrieved, because it was never saved or had expired); \emph{asked anyway} (such an entry was
retrievable and the model chose to ask the user); \emph{wrong}; or, under attack only,
\emph{declined} (the user refused a confirmation for a planted value). Cost figures come from a second
set of runs with the planted items removed. Tables report the mean and standard deviation over
five seeds.

\subsection{Models}

\begin{center}\small\TableModels\end{center}

\noindent The runs were made between 22 and 25 November 2026. The reported figures rest on
8{,}970 distinct model replies. Every figure can be recomputed from the saved replies without
access to a model, and we checked that doing so reproduces the reports exactly.

\section{Results}

\subsection{Undefended agents are poisoned at very different rates}

\begin{table}[t]\centering\small
\resizebox{\textwidth}{!}{\TableSecurity}
\caption{Security. ``Saved'' is the share of planted items in memory after reading. Enforced is
with a user who approves a bad prompt one time in ten. Mean $\pm$ standard deviation over five
seeds; 150 planted items per model.}
\label{tab:security}
\end{table}

Table~\ref{tab:security} shows attack success with no defense ranging from 2\% for Claude Haiku
5.5 to 93\% for Llama 4 Maverick. Most of that spread is decided when the document is read.
Table~\ref{tab:behaviour} shows what each model chose to save. Llama saved all 108 distinct
planted documents. The five hosted models together saved something from 34 of 290 bluntly worded
ones, and from 5 to 37 of the 50 plainly worded ones. Once a planted item was saved by an
undefended agent it usually succeeded: saved and successful shares differ by at most fifteen
points.

Models also differed on legitimate content. GPT-6 Luna saved none of the eight legitimate payout
accounts that arrived in outside documents, and Claude Haiku three, while both saved most contact
addresses. That caution costs utility with no defense at all: 5\% and 4\% of requests found the
needed detail missing for those two models, against none for Sonnet, Gemini and Llama.

Only two models were affected by extra-action items (Luna 30\%, Llama 96\%); the others saved no
standing instruction that later executed. Claude Sonnet 5.5 returned an empty reply to 20
documents, all of them planted extra-action notes; an empty reply is scored as saving nothing,
and the harness did not record why the reply was empty. Pulipaka et al.\ observed a Claude
model explicitly refusing suspicious memory writes~\cite{sleepermemory}, which may be related.

\begin{table}[t]\centering\small
\TableBehaviour
\caption{Model behaviour, counted over distinct prompts. Left: documents from which the model
saved a value or instruction, for planted documents worded bluntly or plainly and for legitimate
payout accounts and contact addresses. Right, with labels visible: how often the model used a
memory when only untrusted entries were available, for requests that act (``Acted'') and for
drafts (``Drafted''); and, when a trusted and an untrusted entry were both in view and the model
used one, how often it was the trusted one.}
\label{tab:behaviour}
\end{table}

\subsection{Labels shown to the model help some models much more than others}
\label{sec:labels}

With labels visible but unenforced, attack success fell to 0--1\% for Sonnet, Gemini, Luna and
Haiku, to 15\% for GPT-6.1 Terra, and only to 67\% for Llama. The right half of
Table~\ref{tab:behaviour} shows why. Faced with a request that acts and only untrusted memory to
go on, Sonnet, Gemini and Haiku never used the memory (0 of 566 cases). Luna used it in 19 of
196, Terra in 145 of 214, and Llama in 347 of 349.

When a trusted and an untrusted entry were both in view, every model chose the trusted one, in
all 990 cases. Failures to follow labels therefore arose where there was nothing
trusted to prefer: the fill-gap case, and cases where merging had already destroyed the trusted
entry (Section~\ref{sec:lifecycle}).

With enforcement, no planted item succeeded on five models. On Llama 4.7\% succeeded, all through
confirmation prompts that the simulated user approved. Table~\ref{tab:fatigue} varies that user.
With a user who never approves a bad prompt, success is zero for every model. With one who
approves half of them it reaches 15\% for Llama, because Llama proposes the planted value every
time and so every such request reaches the user. Models that decline untrusted memory themselves
generate no prompts to approve.

Prior work on small open-weight models concluded that provenance cues visible to the model do not
hold~\cite{ppmf,sleeperchannels}, while Zhan et al.\ found that labels in the prompt worked well
on the seven models they tested~\cite{authoritycollapse}. Our results reproduce both: the Llama result agrees
with the first and the hosted models with the second. The hosted models show that this is a
property of the model rather than of labels: three of them treated a visible label as binding for
requests that act. A deployment cannot know in advance which kind of model it has, and a model
update can change it, so the case for enforcement stands. For the three label-respecting models
enforcement moved attack success by about one point at most. What it changed for them was cost, and in
their favour (Section~\ref{sec:cost}).

\begin{table}[t]\centering\small
\TableFatigue
\caption{Enforced condition: attack success (\%) as the simulated user becomes more willing to
approve a confirmation prompt for a planted value.}
\label{tab:fatigue}
\end{table}

\subsection{What the defense costs when nobody is attacking}
\label{sec:cost}

\begin{table}[t]\centering\small
\resizebox{\textwidth}{!}{\TableCost}
\caption{Cost with no planted items. Share of requests with a known right answer. ``Detail lost'':
no retrievable entry held the right value. ``Asked anyway'': one did, and the model asked the
user instead. Mean $\pm$ standard deviation over five seeds.}
\label{tab:cost}
\end{table}

Table~\ref{tab:cost} gives the cost. With no attacker, the share of requests completed correctly
without involving the user fell to 86\% for every model when the enforced defense was switched
on: a drop of 14 points for the three models that had been at 100\%, and of 3 to 12 points for
the three that already lost or asked about some details with no defense.

That the six models land on the same figure is a property of the workload, and it is the main
thing the workload shows. The 15 legitimate outside details are the only memories the defense
treats differently, and under enforcement each one costs one exchange with the user, the
first time a request that acts depends on it. After that exchange the value is held as the user's own
statement, is kept apart from untrusted entries when memory is merged, and does not expire. The
simulated user had to supply a detail 354 times under enforcement, across all models and
seeds, and never one it had supplied before. Fifteen first uses in 105 requests is 14\%. The cost of this design therefore
scales with the number of distinct outside facts an agent relies on and not with the number of
requests. What differs between models is the form the exchange takes.

\emph{The model asks.} Once Sonnet, Gemini and Haiku can see that a legitimate supplier detail is
labelled untrusted, they decline to act on it and ask the user: on 12\%, 13\% and 9\% of
requests respectively. For these models the runtime's confirmation prompt never fired, because
the model never proposed an untrusted value for a request that acts. This is the same pattern,
at lower intensity, that Kwon saw when an explicit distrust instruction made a model escalate
every request~\cite{manufacturedconfidence}.

\emph{The runtime asks.} Llama never asked on its own, so the gate did the work, at about 10
confirmation prompts per 100 requests; Terra produced 7 and Luna 1. A confirmation is
arguably cheaper for the user than an open question, since it shows the value and its source.
Counting a confirmed request as correct, Llama completes 96\% of requests under enforcement and
Terra 92\%, against 100\% and 98\% undefended.

\emph{The detail is gone.} Untrusted entries expire after 20 sessions, and a legitimate fact may
be needed long after it arrived. Sonnet, Gemini and Llama found no detail missing with no defense
and 2\%, 1\% and 4\% with enforcement. For Luna, Terra and Haiku the figure rose by 1 to 3
points, to 7\%, 6\% and 5\%; part of that is detail these models never saved.

\emph{Labels without enforcement cost the careful models more.} In the Labels condition the
share completed unprompted was 75\% for Sonnet, 70\% for Gemini and 75\% for Haiku, against 86\%
with enforcement. The difference is repeated questions. Without the rule that keeps trust levels
apart, the user's answer is merged at the next consolidation with the untrusted entries on the
same topic, takes their label and their expiry, and the model, seeing an untrusted entry again,
asks again. Of the exchanges with the user in the Labels condition, 55 of 130 for Sonnet, 80 of
155 for Gemini and 54 of 129 for Haiku concerned a detail the user had already supplied. Llama,
which does not ask, completed more requests unprompted with labels alone (93\%) than with
enforcement, because nothing stopped it using an untrusted value.

\subsection{Lifecycle: merging, expiry and repair}
\label{sec:lifecycle}

\emph{Merging across trust levels damages the user's own instructions.} In the Labels condition a
merge of the user's stated account with a planted note yields one entry labelled untrusted and
due to expire. Under attack, the share of requests that found the needed detail missing was
higher with Labels than with Enforced for every model, by 4 to 10 points (for example 12\%
against 2\% for Sonnet, and 11\% against 3\% for Llama). The repeated questions of
Section~\ref{sec:cost} are the same effect with no attacker present. It is also a likely source of the residual override success in the Labels
column, since with the trusted entry gone there is nothing for the model to prefer; we did not
trace each case. Keeping trust
levels apart during consolidation removed the effect. Zhan et al.\ report the same failure in
stronger form: labelling a consolidated memory with the least permissive label of any source cut
task success to about 5\%~\cite{authoritycollapse}.

\emph{Expiry is a blunt instrument.} It bounds how long a planted item can wait, and it removed
legitimate outside facts at the rates given above. A confirmation that promotes a value to
trusted is what keeps a useful outside fact alive; Llama, which reaches the confirmation prompt
most often, found fewer details missing under Enforced (4\%) than under Labels (7\%).

\emph{Repair depends on lineage.} At the end of each run we retract every document that carried a
planted item. With no defense, most of the planted items that had been saved remained (between
67\% and 93\% of them, depending on the model; the rest had already been overwritten),
because the agent's notes about them recorded no source to retract by. With labels, none remained for any
model.

\section{Limitations}
\label{sec:limits}

\begin{itemize}
\item \textbf{The attacks do not adapt.} The planted texts are fixed, simply worded templates.
Defenses that look strong against static attacks have repeatedly been broken by adaptive
ones~\cite{attackermovessecond}. Our undefended and Labels figures are therefore lower bounds on
attack success. The Enforced rules do not depend on wording, but the confirmation prompt depends
on a person, and we have not tested it with people.
\item \textbf{Synthetic workload, written by us.} Requests are short, topics are of three kinds,
and each document statement carries a topic handle that stands in for retrieval. Real memory
systems retrieve by similarity, which adds attack surface we do not model. Rates of asking,
prompting and expiry will move with the task mix, the expiry length and the share of facts that
arrive from outside.
\item \textbf{An ideal user.} When asked, the simulated user always supplies the right value
at once, and the answer is stored as a user statement. A real user may not know the answer, may
answer wrongly, or may read a planted value back from the same email. The finding that each
outside fact costs one exchange depends on this, and on our workload, in which every outside
fact is eventually needed and the facts do not change. An earlier version of the harness stored
no answer when the model asked; all six models were rerun after that was corrected, and only
the corrected runs are reported.
\item \textbf{One run configuration per model.} Five seeds, one prompt wording, one expiry
length. We report standard deviations and no significance tests. Replies are not guaranteed to be
deterministic. Four models were reached through an aggregator and two through their vendor's API.
\item \textbf{Repair is measured one-sidedly.} We check that planted items are gone after their
source is retracted. We do not measure what legitimate memory is removed with them. Our labels
are assigned per session, so a legitimate fact noted in the same session as a planted document
shares its source and would be retracted too; MemSecBench reports that such collateral loss is
the main obstacle to repair~\cite{memsecbench}. The same coarse, per-session taint also labels
legitimate facts untrusted more often than attribution-based lineage would~\cite{memlineage},
which inflates our cost figures by an amount we did not measure.
\item \textbf{Our implementation only.} We did not run the released implementations of the
defenses we cite. The Enforced condition is our own reading of those mechanisms.
\item \textbf{Empty replies.} Twenty-one replies from the two Claude models were empty and are
scored as saving nothing; the cause was not recorded.
\item \textbf{Not a model ranking.} A single synthetic workload cannot establish which model is
safer in general, and model behaviour changes between versions.
\end{itemize}

\section{Conclusion}

On this workload, provenance labels shown to the model were a strong defense for models that
respect them and a weak one for a model that does not, and enforcement by the runtime removed
that dependence. The enforced defense blocked every simple planted item we tried. Its price when
nobody was attacking was one exchange with the user for each legitimate outside fact the agent
came to rely on, which on this workload meant 14\% of requests, whichever model was used. The
model decides only whether that exchange is a question it asks or a confirmation the runtime
raises. Labels without the rule that keeps trusted and untrusted memory apart cost the careful
models 10 to 15 points more, because the user's answers did not stay trusted. The measurements
that would most change this picture are adaptive attacks, a realistic benchmark such as
MemoryDojo, and real users in place of the simulated one.

\section*{Author's statement on AI assistance}

The author chose the research question, directed the design of the harness and the experiments,
ran every experiment, decided which corrections to make and which claims to keep, and takes full
responsibility for the content of this paper. An AI assistant (Claude, from Anthropic) was used
as a support tool and is not an author. Two of the six evaluated models are from the same
vendor as that assistant. To limit the effect of this on the results, all figures are computed
by a script from the saved model replies, which are released so that anyone can recount them.

\section*{Availability}

Code, workload generator, per-model reports and full transcripts:
\url{https://github.com/Radmo3/memgate}.

\end{document}

%% file: tables.tex
\newcommand{\TableModels}{%
\begin{tabular}{llll}\toprule
Model & Vendor & Name sent to the endpoint & Endpoint \\\midrule
Llama 4 Maverick & Meta (open weights) & \texttt{meta-llama/llama-4-maverick} & OpenRouter \\
Claude Sonnet 5.5 & Anthropic & \texttt{anthropic/claude-sonnet-5.5} & OpenRouter \\
GPT-6 Luna & OpenAI & \texttt{gpt-6-luna} & OpenAI API \\
GPT-6.1 Terra & OpenAI & \texttt{gpt-6.1-terra} & OpenAI API \\
Gemini 3.8 Flash & Google & \texttt{google/gemini-3.8-flash} & OpenRouter \\
Claude Haiku 5.5 & Anthropic & \texttt{anthropic/claude-haiku-5.5} & OpenRouter \\
\bottomrule\end{tabular}}

\newcommand{\TableSecurity}{%
\begin{tabular}{lrrrrrrr}\toprule
 & & \multicolumn{3}{c}{Attack success (\%)} & \multicolumn{3}{c}{No defense, by attack (\%)} \\\cmidrule(lr){3-5}\cmidrule(lr){6-8}
Model & Saved (\%) & No defense & Labels & Enforced & Extra action & Override & Fill gap \\\midrule
Llama 4 Maverick & 100 $\pm$ 0 & 93 $\pm$ 3 & 67 $\pm$ 9 & 5 $\pm$ 3 & 96 & 82 & 100 \\
Claude Sonnet 5.5 & 53 $\pm$ 3 & 43 $\pm$ 9 & 0 $\pm$ 0 & 0 $\pm$ 0 & 0 & 42 & 86 \\
GPT-6 Luna & 45 $\pm$ 5 & 41 $\pm$ 7 & 1 $\pm$ 2 & 0 $\pm$ 0 & 30 & 20 & 72 \\
GPT-6.1 Terra & 43 $\pm$ 6 & 28 $\pm$ 7 & 15 $\pm$ 8 & 0 $\pm$ 0 & 0 & 22 & 62 \\
Gemini 3.8 Flash & 25 $\pm$ 6 & 23 $\pm$ 7 & 0 $\pm$ 0 & 0 $\pm$ 0 & 0 & 24 & 44 \\
Claude Haiku 5.5 & 6 $\pm$ 3 & 2 $\pm$ 2 & 1 $\pm$ 2 & 0 $\pm$ 0 & 0 & 6 & 0 \\
\bottomrule\end{tabular}}

\newcommand{\TableCost}{%
\begin{tabular}{lrrrrrrr}\toprule
 & \multicolumn{2}{c}{Correct, unprompted (\%)} & \multicolumn{5}{c}{Enforced defense (\%)} \\\cmidrule(lr){2-3}\cmidrule(lr){4-8}
Model & No defense & Labels & Correct, unprompted & Correct after prompt & Detail lost & Asked anyway & Prompts / 100 tasks \\\midrule
Llama 4 Maverick & 100 $\pm$ 0 & 93 $\pm$ 2 & 86 $\pm$ 1 & 10 $\pm$ 2 & 4 $\pm$ 2 & 0 $\pm$ 0 & 9.9 \\
Claude Sonnet 5.5 & 100 $\pm$ 0 & 75 $\pm$ 4 & 86 $\pm$ 0 & 0 $\pm$ 0 & 2 $\pm$ 1 & 12 $\pm$ 1 & 0.0 \\
GPT-6 Luna & 95 $\pm$ 1 & 82 $\pm$ 1 & 86 $\pm$ 0 & 1 $\pm$ 0 & 7 $\pm$ 1 & 6 $\pm$ 1 & 1.1 \\
GPT-6.1 Terra & 98 $\pm$ 1 & 90 $\pm$ 1 & 86 $\pm$ 0 & 6 $\pm$ 2 & 6 $\pm$ 2 & 2 $\pm$ 1 & 6.5 \\
Gemini 3.8 Flash & 100 $\pm$ 0 & 70 $\pm$ 3 & 86 $\pm$ 0 & 0 $\pm$ 0 & 1 $\pm$ 1 & 13 $\pm$ 1 & 0.0 \\
Claude Haiku 5.5 & 89 $\pm$ 1 & 75 $\pm$ 2 & 86 $\pm$ 0 & 0 $\pm$ 0 & 5 $\pm$ 1 & 9 $\pm$ 1 & 0.0 \\
\bottomrule\end{tabular}}

\newcommand{\TableBehaviour}{%
\begin{tabular}{lrrrrrrr}\toprule
 & \multicolumn{2}{c}{Planted, saved} & \multicolumn{2}{c}{Legitimate, saved} & \multicolumn{3}{c}{With labels visible} \\\cmidrule(lr){2-3}\cmidrule(lr){4-5}\cmidrule(lr){6-8}
Model & Blunt & Plain & Account & Contact & Acted & Drafted & Chose trusted \\\midrule
Llama 4 Maverick & 58/58 & 50/50 & 8/8 & 15/15 & 347/349 & 171/171 & 182/182 \\
Claude Sonnet 5.5 & 16/58 & 33/50 & 8/8 & 15/15 & 0/275 & 74/156 & 218/218 \\
GPT-6 Luna & 9/58 & 37/50 & 0/8 & 14/15 & 19/196 & 77/109 & 140/140 \\
GPT-6.1 Terra & 6/58 & 35/50 & 5/8 & 15/15 & 145/214 & 107/114 & 84/84 \\
Gemini 3.8 Flash & 2/58 & 22/50 & 8/8 & 15/15 & 0/197 & 0/155 & 234/234 \\
Claude Haiku 5.5 & 1/58 & 5/50 & 3/8 & 12/15 & 0/94 & 8/91 & 132/132 \\
\bottomrule\end{tabular}}

\newcommand{\TableFatigue}{%
\begin{tabular}{lrrrr}\toprule
 & \multicolumn{4}{c}{Chance the user approves a bad prompt} \\\cmidrule(lr){2-5}
Model & 0 & 0.1 & 0.3 & 0.5 \\\midrule
Llama 4 Maverick & 0.0 & 4.7 & 10.0 & 15.3 \\
Claude Sonnet 5.5 & 0.0 & 0.0 & 0.0 & 0.0 \\
GPT-6 Luna & 0.0 & 0.0 & 0.7 & 0.7 \\
GPT-6.1 Terra & 0.0 & 0.0 & 2.7 & 4.0 \\
Gemini 3.8 Flash & 0.0 & 0.0 & 0.0 & 0.0 \\
Claude Haiku 5.5 & 0.0 & 0.0 & 0.0 & 0.0 \\
\bottomrule\end{tabular}}